\documentclass[dvips]{acta}
\usepackage{amssymb}
\usepackage{amsmath}
\usepackage{ctable}
\SetPages{0}{0}
\SetVol{61}{2011}

\long\def\symbolfootnote[#1]#2{\begingroup
\def\thefootnote{\fnsymbol{footnote}}\footnote[#1]{#2}\endgroup} 

\begin{document}

 \begin{Titlepage}
\Title{Photometric study of the star cluster NGC~2155 \\ in the Large~Magellanic~Cloud: age estimation and variable stars.\symbolfootnote[1]{This paper includes data gathered with the 6.5 meter Magellan Telescopes
located at the Las Campanas Observatory, Chile.}}

\Author{M.~~O~t~u~l~a~k~o~w~s~k~a$^1$,~~A.~~O~l~e~c~h$^1$,~~W.~~P~y~c~h$^1$, \\ ~~A.~A.~~P~a~m~y~a~t~n~y~k~h$^{1,2}$,~~T.~~Z~d~r~a~v~k~o~v$^1$,~~S.~~M.~~R~u~c~i~n~s~k~i$^3$}
{$^1$N. Copernicus Astronomical Center, Polish Academy of Sciences, ul. Bartycka 18, 00-716 Warsaw, Poland \\
e-mail: magdaot@camk.edu.pl \\
$^2$Institute of Astronomy, Russian Academy of Sciences, Pyatnitskaya Str. 48, 109017 Moscow, Russia \\
$^3$Department of Astronomy and Astrophysics,
University of Toronto, Toronto, \\ ON M5S 3H4, Canada }

\Received{Month Day, Year}
\end{Titlepage}

\Abstract{
We present results of new photometry for the globular star cluster NGC~2155 in the Large Magellanic Cloud (LMC). 
Our \textit{I-} and \textit{V-}band observations were obtained with the 6.5-meter Magellan~1 Baade Telescope at Las Campanas Observatory resulting in deep photometry down to $V \sim 24$~mag.
By analyzing the color -- magnitude diagram 
for the cluster and utilizing the 
\textit{Victoria-Regina} grid of isochrones models
we estimated the age of the cluster at $\simeq 2.25$ Gyr
and [Fe/H]=$-0.71$, the numbers which place 
NGC~2155 outside the age-gap in the age-metallicity relation for LMC clusters. 
Using the \textit{Difference Image Analysis Package (DIAPL)},
we detected 7 variable stars in the cluster field
with variability at the level of 0.01~magnitude in the \textit{I}-band. 
Three variables are particularly interesting: 
two SX~Phoenicis (SX~Phe) stars pulsating in the fundamental mode, 
and a detached eclipsing binary which is a prime candidate 
to estimate the distance to the cluster.
}{methods: observational - techniques: photometric - galaxies: individual: LMC - Magellanic Clouds - galaxies: star clusters - stars: distances - stars: variables - globular clusters: individual: NGC~2155 - color-magnitude diagrams}

\section{Introduction}
\label{int}

The metallicity distribution and star formation history of the Large Magellanic Cloud (LMC) has been the subject of many recent studies, for instance Cole~et~al. (2009).
One of the most intriguing problems is the formation and spatial distribution of the LMC star clusters. 
This topic was recently studied for example by Beasley, Hoyle, and Sharples~(2002) and~Baume~et~al. (2007). 

A very peculiar issue about evolution of clusters in this galaxy 
is the existence of so-called "age gap" in the relation between their age and metallicity. 
This fact was mentioned for the first time by Jensen, Mould, and~Reid~(1988), who found that none of LMC clusters has the age between 4 and 10~Gyr. 
Many further studies were carried out to assign the properties of the "age gap", for instance 
Geisler~et~al. (1997) studied 25 LMC clusters and found no cluster formation during the period 3-8~Gyr ago. They reported the cluster formation to start again about 3~Gyr ago with a peak at about 1.5~Gyr ago.
Sarajedini (1998) discovered three star clusters with the age of about 4~Gyr, thus in the "age gap", among them NGC~2155. Following studies showed these ages were perhaps overestimated, for example Kerber, Santiago, and Brocato (2007), and the series of three publications based on the VLT data: Gallart~et~al. (2003), Woo~et~al. (2003), Bertelli~et~al. (2003).
The most recent determinations show the range of the "age-gap" to be about 3 - 10 Gyrs, see Balbinot~et~al. (2010). 
The corresponding metallicity gap, which is less discussed in the literature, is also very evident for the
LMC clusters. It ranges from [Fe/H]$\approx-0.7$, for younger LMC clusters, to [Fe/H]$\approx-2.0$ for older ones, see Olszewski~et~al. (1991).
Nevertheless, the problem of the "age gap" in the age-metallicity relation for LMC clusters still remains unsolved. The NGC~2155 cluster is one of the candidates to fit in the gap.
In this contribution we report our estimate of the age of NGC~2155.
We also address the matter of
variable stars in the field of NGC~2155; currently none has been known. 

The paper is arranged as follows.
In section \ref{obs} we give information on the observations and data reduction.
Color-Magnitude Diagram for the NGC~2155 is presented in section \ref{cmd}
In section \ref{age} we describe isochrones fitting and age estimation for the cluster.
Section \ref{var} is dedicated to variable stars which we found in the field of NGC~2155.
In section \ref{sxphe} we discuss the most interesting variables we found, two SX~Phe stars.
We summarize the conclusions of this work in section \ref{sum}

\section{Observations and data reduction}
\label{obs}

We conducted photometric observations
of the star cluster NGC~2155 in the Large Magellanic Cloud (LMC)
on four nights between 07 December 2002 and 11 December 2002.
We used the TEK\#5 CCD camera attached to the Baade $6.5$~m telescope
(Magellan~1) at the Las Campanas Observatory, Chile.
The field of view was 2.36~x~2.36~arcminutes with the scale
0.069~arcsec/pixel. The
Johnson-Cousins \textit{B}, \textit{V} and \textit{I} passband filters 
were used. The
exposure times for \textit{B} were set 
at  300 and 600 seconds while
the exposure times for \textit{V} and \textit{I} were set 
at 300 seconds.

On the night of 
07/08 December 2002, due to poor weather conditions,
we managed to obtain only 11 images
of NGC2155 in \textit{I} and 3 images in \textit{V}.
These images suffered from poor seeing and have not been used in our
analysis.
On the night of 08/09 December 2002, 
we obtained 48 images in \textit{I}, 12 images
in \textit{V} and 2 images in \textit{B};
we have selected only 14 frames in \textit{I} for further analysis.
On the night of 09/10 December 2002, 
we obtained 50 images in \textit{I}, 14 images
in \textit{V} and 3 images in \textit{B}; we have selected 
45 frames in \textit{I}
and 8 frames in \textit{V} for further analysis. 
Finally, on the night of 10/11 December 2002, 
we obtained 55 images in \textit{I}, 11 images
in \textit{V} and 1 image in \textit{B} of which, for 
further processing we used 47 frames in \textit{I} 
and 9 frames in \textit{V}. A summary of the number of
used frames is given in the Table~\ref{tab:obs}
For the used 106 frames in \textit{I} and 17 frames in \textit{V},
the median seeing was 0.88~arcsec and 0.94~arcsec~(FWHM), respectively.

\begin{table}[h]\footnotesize
\begin{center}
\caption{Log of observations of NGC~2155 for the nights 7/8$-$10/11 Dec 2002}
\begin{tabular}{  c | c | c | c | c }
\hline
Filter	& Exposure time	& Number of  		& Number of			& Median seeing	of selected \\
		& [sec]			& obtained frames	& selected frames 	& frames [arcsec]	\\
\hline
\textit{B}	 &	600.0		&     6			&     0 	&         - 		\\
\textit{V}	 &	300.0		&     40  		&     17 	&         0.94 		\\
\textit{I}  &	300.0		&     164  		&     106 	&         0.88		
\end{tabular}
\end{center}
\label{tab:obs}
\end{table}

Bias subtraction and flat-field correction of the images were done
using the standard procedures within \textit{IRAF}\footnote{IRAF is distributed by the National Optical Astronomical
Observatories, operated by the Association of Universities for Research
in Astronomy, Inc., under contact with the National Science Foundation}
{\it noao.imred.ccdred} package.
We have used programs from the \textit{DIAPL}\footnote{\texttt{http://users.camk.edu.pl/pych/DIAPL/}} and \textit{DAOphotII} (Stetson 1987) packages to perform image subtraction and photometry.

\begin{figure}[h]
\begin{center}
\fbox{\includegraphics[width=0.4\textwidth]{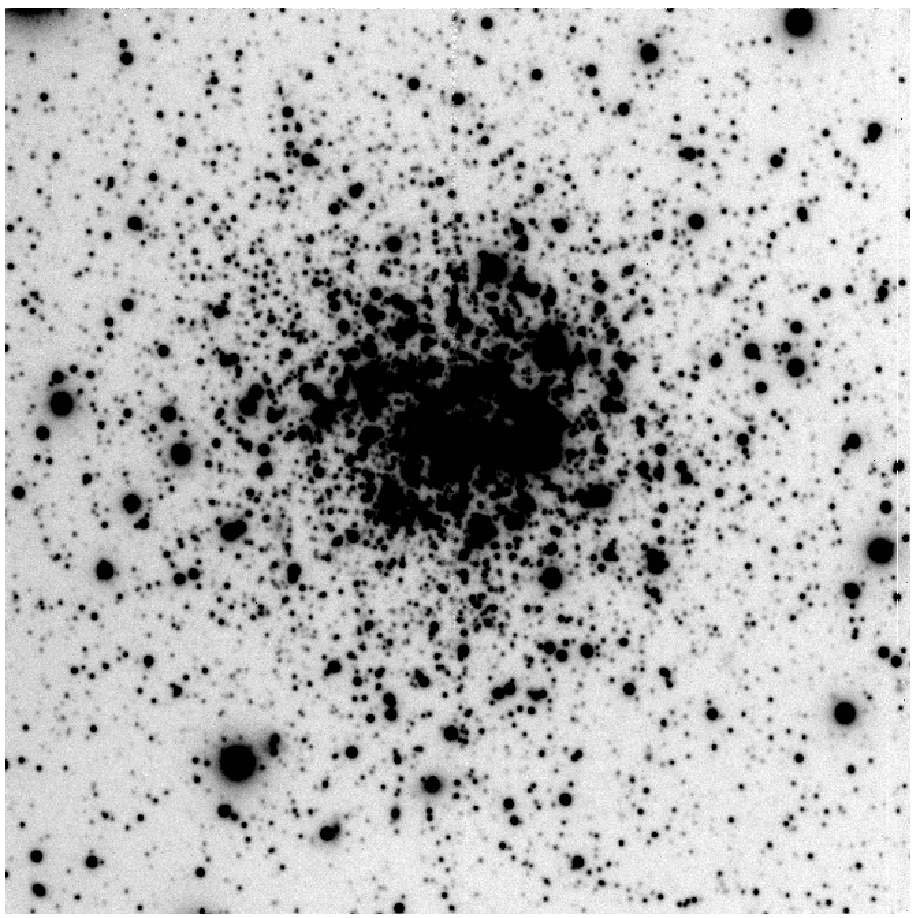}}
\fbox{\includegraphics[width=0.4\textwidth]{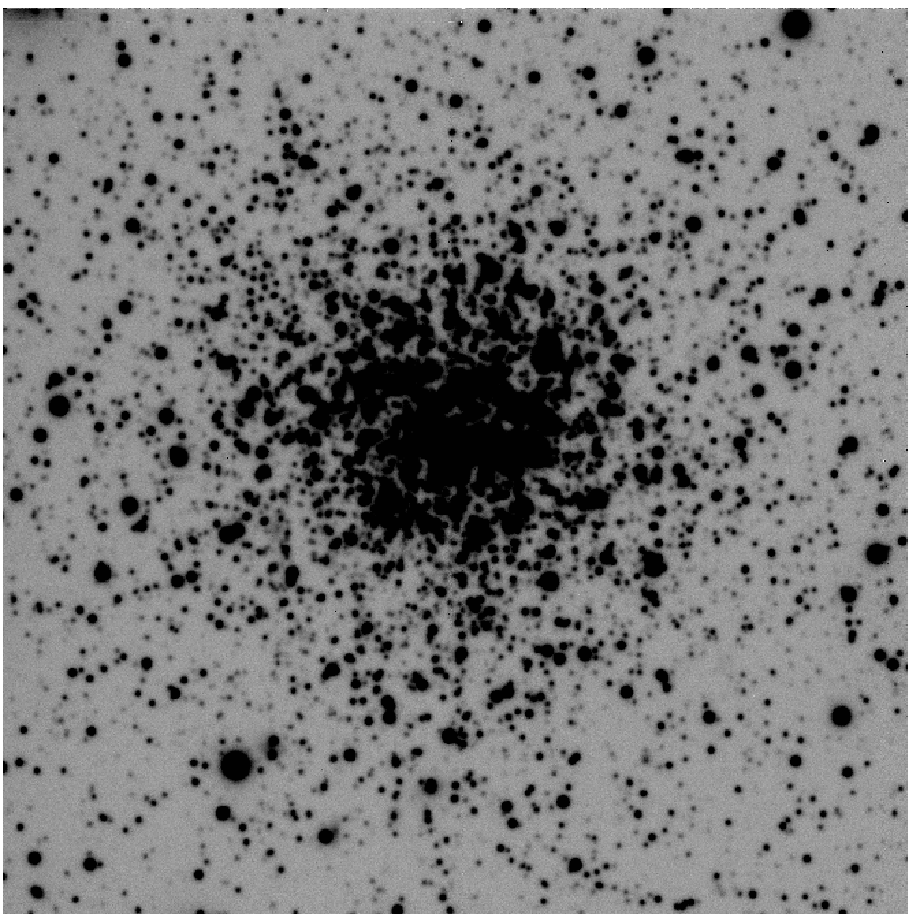}}
\caption{Our "template" images obtained from \textit{DIAPL} package. Left: \textit{I}-band, right: \textit{V}-band.}
\label{fig:templates}
\end{center}
\end{figure}

Using \textit{DIAPL}, we created the 
"template" images of the NGC~2155 
from the best 10 frames in \textit{I} and 8 frames in \textit{V}.
They are shown in the Fig~\ref{fig:templates}. The figure
clearly shows the very highly compact nature of the cluster.

The standard routines of the software package
\textit{DIAPL} allowed us to detect variable stars based on the data from 
only the two last nights of observations which had the acceptable weather conditions. 
The package operates through subtraction
of a template frame from each of the obtained frames.
The output from \textit{DIAPL} was a set of files with 
photometry in arbitrary analog counts. 
After that, 
we converted the light curves from counts to instrumental magnitudes, and later on to standard magnitudes as described below.

Stellar profile photometry was performed using the \textit{DAOPHOT}/\textit{ALLSTAR} routines of \textit{DAOphotII}, see Stetson (1987). 
The PSF model for every frame was obtained from a set of unsaturated and isolated stars. 
We extracted instrumental photometry for all 2715 stars in the template list. 

\begin{figure}[h]
\begin{center}
\includegraphics[width=0.34\textwidth, angle=270]{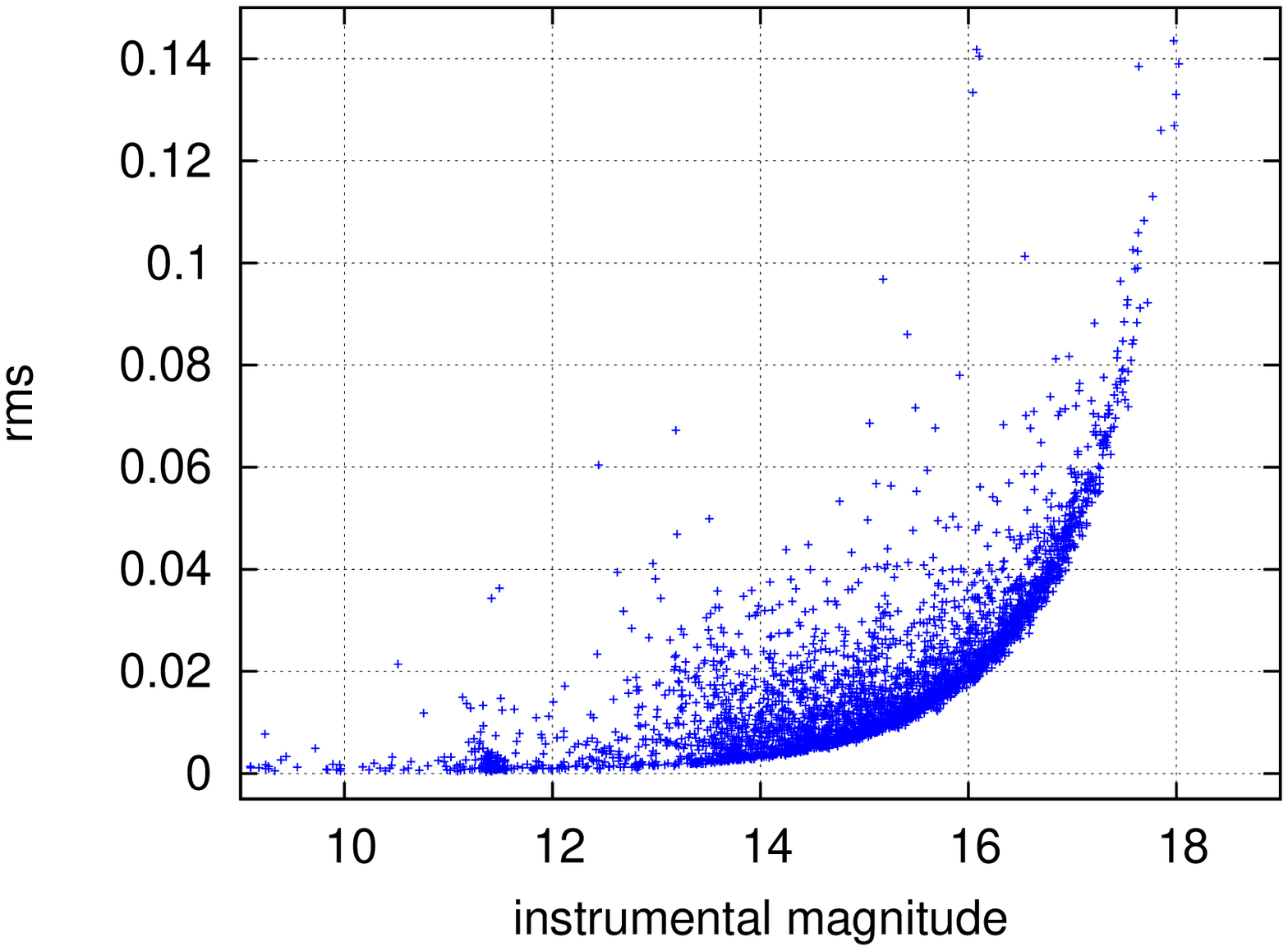}
\includegraphics[width=0.34\textwidth, angle=270]{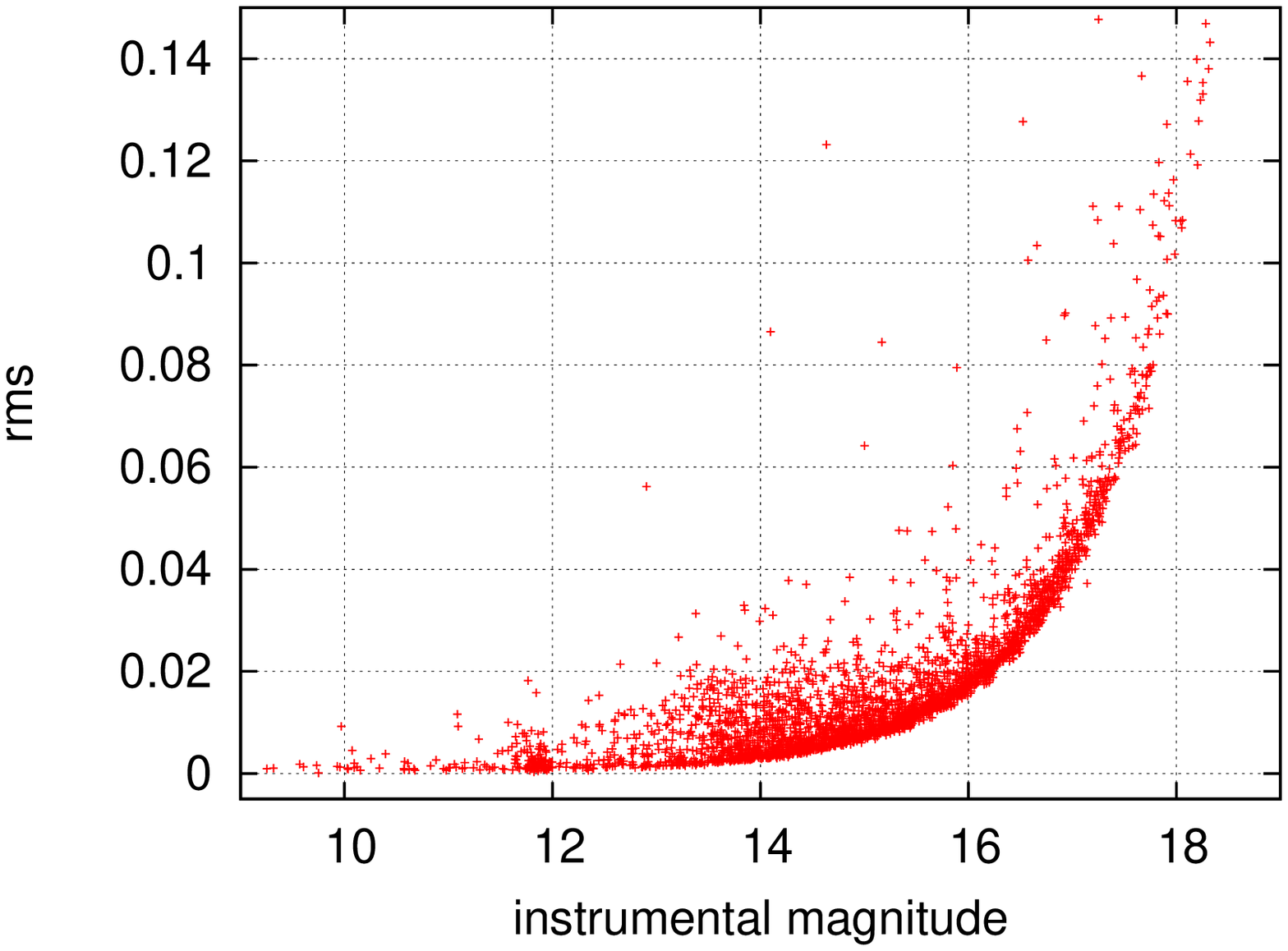}
\caption{A test of the photometry quality: the relation between instrumental magnitude and its {\it rms} deviation obtained from \textit{ALLSTAR}. Left: results for the \textit{I}-band, right: results for the \textit{V}-band.}
\label{fig:mag-sig}
\end{center}
\end{figure}

In the Fig.~\ref{fig:mag-sig}, we present {\it rms} errors as a function of instrumental magnitude for the both template images. 
Positions of the red clump in these plots at
the instrumental magnitudes 
\textit{i} $\approx 11.5$ and
\textit{v} $\approx 12$ 
are clearly visible. 

\subsection{Calibration}

The following equations were used to transform our data to the 
magnitude system of the previous observations
of the cluster by Zaritsky~et~al. (2004):

\begin{equation}
I - i = \alpha_1 + \beta_1 (v - i)
\end{equation}
\begin{equation}
V - v = \alpha_2 + \beta_2 (v - i)
\end{equation}
where $I, V$ refer to standard system values, and $i, v$ are our instrumental magnitudes.

We found:
$ \alpha_1  = 6.320 \pm 0.017 $,
$ \beta_1 = 0.213 \pm 0.028 $,
$ \alpha_2  = 6.819 \pm 0.014 $ and
$ \beta_2 = 0.272 \pm 0.023 $.

\subsection{Color-Magnitude Diagrams (CMD)}
\label{cmd}

\begin{figure}[h]
\begin{center}
\includegraphics[width=0.45\textwidth, angle=270]{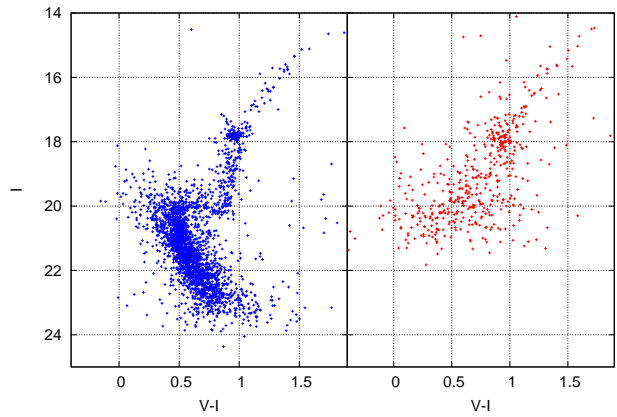}
\caption{CMD comparison. Left: our CMD, right: CMD of Zaritsky~et~al. (2004).}
\label{fig:cmd_compare1}
\end{center}
\end{figure}

We compared our CMD (before extinction corrections) with the CMD based on the data of Zaritsky~et~al. (2004), which has been used as standard magnitudes in our calibration procedure. The diagrams are shown in the Fig.~\ref{fig:cmd_compare1}.
The comparison confirms that our calibration to standard magnitudes went fine: we can see the location of the red clump of our CMD is in the same place as in the data of Zaritsky~et~al. (2004), 
although our photometry appears to be by far more accurate.

\begin{figure}[h]
\begin{center}
\includegraphics[width=0.45\textwidth, angle=270]{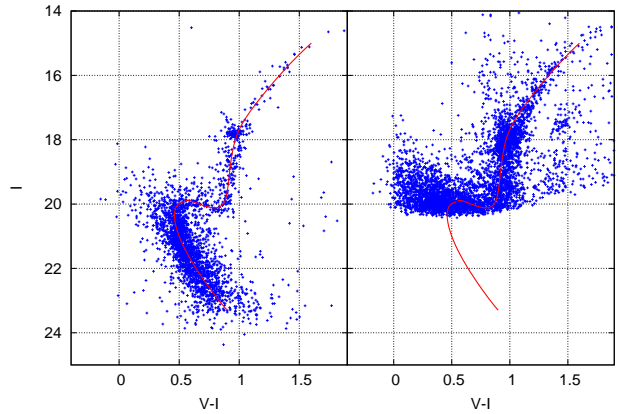}
\caption{CMD comparison. Left: our CMD, right: CMD of Udalski (1998).}
\label{fig:cmd_compare2}
\end{center}
\end{figure}

Another test of our transformations
was a comparison with the data of Udalski~(1998), as shown
in Fig.~\ref{fig:cmd_compare2}.
We added a curve representing the trend of our data points to 
both panels of the figure to facilitate the comparison. 
Again, the results are in agreement and what is more, we can clearly see how much deeper our CMD is. 
This gives us a chance to successfully determine the age and metallicity of the cluster from
the location of its Main Sequence.

\subsection{Extinction correction}
\label{ext}

\begin{figure}[h]
\begin{center}
\includegraphics[width=0.7\textwidth, angle=270]{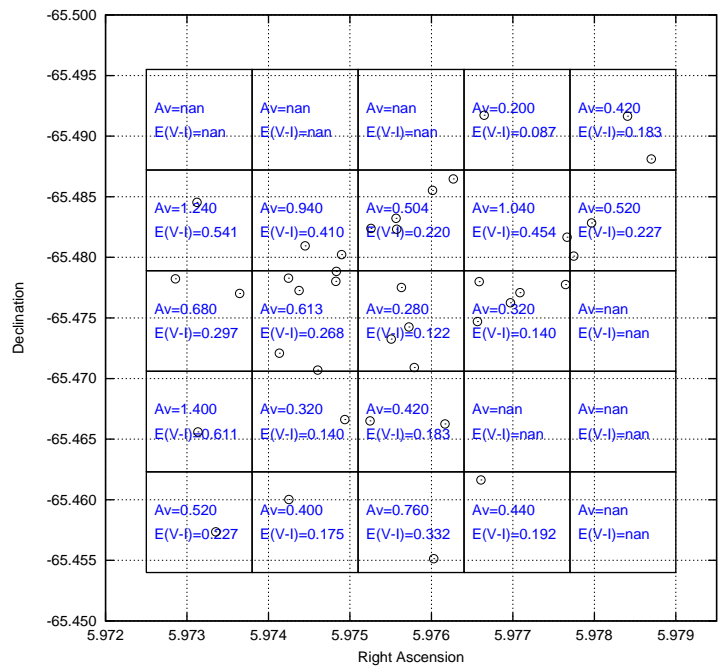}
\caption{An example map of averaged values of interstellar extinction for each subfield. Dots represent data points from Zaritsky~et~al. (2004). 
"nan" stands for "Not a Number" which is a result for subfields with no data points.}
\label{fig:map}
\end{center}
\end{figure}

Originally we intended to apply 
position-dependent extinction corrections within
the field of the cluster, so we divided the field into a mosaic of subfields, as shown in the Fig.~\ref{fig:map}. 
We made tests for such a grid of
 $3\times3$ to $5\times5$, each time calculating the average value of extinction for every subfield 
using values derived with the online tool 
\textit{Reddening Estimator for the LMC} by Zaritsky~et~al. (2004)\footnote{\texttt{http://ngala.as.arizona.edu/dennis/lmcext.html}}.  

However, during this process we realized
that this procedure leads to too few stars per sub-field
to make this kind of statistics reasonable.
Thus, 
we decided to use the mean value of extinction for the whole field $2.5\times2.5$[arcmin], which is the mean value calculated using \textit{all stars model} (hot and cool stars) of Zaritsky~et~al. (2004): $\langle A_V \rangle$~=~0.56, 
centered on right ascension $\alpha=5.97567$[h] and declination $\delta=-65.47722{\rm[^o]}$.

Following Schlegel, Finkbeiner, and Davis (1998) we assumed:
\[ 	\frac{A_V}{E(B-V)} = 3.315 \]
\[ 	\frac{A_I}{E(B-V)} = 1.940 \]
leading to:
\[ \frac{A_V}{E(V-I)} = 2.410 \]

\section{Age estimate for NGC~2155}
\label{age}

There have been several attempts to estimate the age of NGC~2155; the most recent are listed in the~Tab.~2. 
The results differ because of variety of combinations of the used parameters and isochrone models. 
The range of the derived ages is 2.3 to 3.6 Gyr, with most estimates falling within 2.5 -- 3.0 Gyr.

\begin{footnotesize}
\ctable[
   caption = Recent age estimations for LMC star cluster NGC~2155. \textit{Z} is the mass fraction abundances of all elements heavier than helium. Standard errors are given when available. Isochrones names are clarified in the given references.
]{ccccccc}{
\tnote[a]{Rich, Shara and Zurek (2001)}
\tnote[b]{Piatti et al. (2002)}
\tnote[c]{Gallart et al. (2003)}
\tnote[d]{Bertelli et al. (2003)}
\tnote[e]{Woo et al. (2003)}
\tnote[f]{VandenBerg, Bergbusch, and Dowler (2006)}
\tnote[g]{Kerber, Santiago, and Brocato (2007)}
\tnote[h]{This work}
} {	\FL
theoretical  	& distance 				& filters 	&	metallicity	&	Z				&  obtained			& source	\NN
 isochrone 		& modulus				& used		&	[Fe/H]		&					&  age 				& 			\NN
 name			& $(m-M)_V$				&			&				&					&  [Gyr] 			& 			\ML
 Girardi			&  					& $B, V$	& $-0.68$		&					&  3.2				& \tmark[a]	\NN	
 Geneva				& 18.55$\pm$0.10	& $C, T_1$	& $-0.9\pm0.2$ 	& 0.004				&  3.6$\pm$0.7		& \tmark[b]	\NN
 Yonsei-Yale		& 18.5				& $V, R$	& $-0.7	$		& 0.004				&  3.0				& \tmark[c] \NN
 Padova				& 18.5				& $V, R$	& $-0.7 $		& 0.004				&  2.5				& \tmark[c] \NN
 Padova				&  					& $V, R$	& 				& 0.003				&  2.8				& \tmark[d] \NN
 Yonsei-Yale		& 18.5				& $V, R$	& 				& 0.004				&  2.7				& \tmark[e] \NN
 Yonsei-Yale		& 18.5				& $V, R$	& 				& 0.004				&  2.9				& \tmark[e] \NN	
 Victoria-Regina 	& 18.65		 		& $V, R$	& $-0.83$		&					&  2.3				& \tmark[f] \NN
 - 					& 18.32$\pm$0.04	& $B, V$	& $-0.7\pm0.1$	& 0.004$\pm$0.001	&  3.0$\pm$0.25		& \tmark[g] \NN
 Victoria-Regina  	& 18.5				& $V, I$ 	& $-0.71$		& 					&  2.25				& \tmark[h] \LL
}
\end{footnotesize}

To estimate the age of NGC~2155 based on our CMD, we used the isochrones called \textit{The Victoria-Regina Stellar Models} from VandenBerg, Bergbusch, and Dowler (2006).
From the grid of evolutionary tracks we 
chose the one which is closest to our CMD in its shape. 
We tried a wide range of metallicities and ages with, and without, the convective-core overshooting. We found the best isochrone for our CMD is the one with the metallicity of [Fe/H]~=~$-0.71$, age~2.25~Gyr, and with the overshooting.

There are a few parameters which have an influence on the shape of isochrones. They are the main sources of uncertainties while estimating the age of a star cluster. Especially 
assumptions on the distance to the cluster and the chosen metallicity
are the reasons for cluster age differences between the studies.  

\begin{figure}[h]
\begin{center}
\includegraphics[width=0.75\textwidth, angle=270]{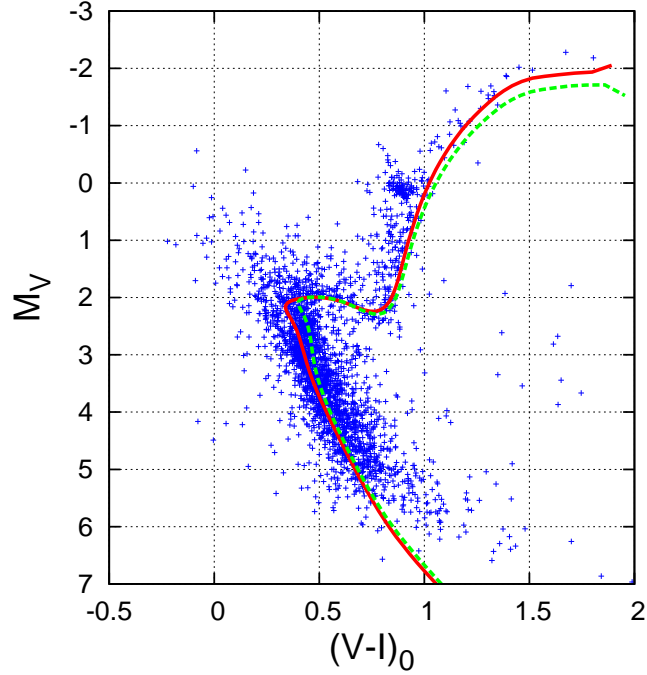} 
\caption{Our best isochrones fitting: two isochrones of the metallicity [Fe/H]~=~$-0.71$ (green dashed curve) and [Fe/H]~=~$-0.83$ (red solid curve), the age of 2.25~Gyr, and with convective core overshooting, are plotted on our CMD. See description in text.}
\label{fig:cmd_iso}
\end{center}
\end{figure}

For the bolometric correction we applied the values from Bessell, Castelli, and Plez (1998), and for the interstellar reddening the values from Schlegel, Finkbeiner, and Davis (1998). For the value of the distance modulus to the LMC, we have chosen $m - M = 18.50 \pm 0.13$, following Benedict~et~al. (2002).
We used studies from the~Tab.~2. and references in there to find the range of metallicity for NGC~2155. Considering this data we fitted isochrones of each metallicity possible to find the best one (Fig.~\ref{fig:cmd_iso}).
A very influential parameter for the estimate of the age of the cluster is the convective core overshooting. As was shown for instance by Woo~et~al. (2003), it has a great impact on the shape of the CMD and must be provided for intermediate-age star clusters. 
The effect of the existence of binaries and field stars on the CMD has a minor influence on the age estimate and is difficult to calculate. Thus, we do not consider it in our study, as was also done in most of prior studies. Although taking into consideration the binary fraction could improve the fitting of isochrones along the main sequence.

In the Fig.~\ref{fig:cmd_iso}, the green (dashed) curve gives our best isochrone fitting to the CMD of NGC~2155.
The choice of the best isochrone is quite ambiguous, though. In this figure we can see a comparison of two isochrones with the age of 2.25~Gry of metallicity [Fe/H]~=~$-0.71$ (the green dashed curve) and [Fe/H]~=~$-0.83$ (the red solid one). 
Along the main-sequence (MS) the green curve fits the CMD better than the red one: it is located in the very center of the MS, whereas the red curve is shifted to the left.
The opposite situation appears on the upper part of the CMD: the trend of our CMD is better represented by the red curve.
However, we need to remember that we used a grid of isochrones, thus the real metallicity of NGC~2155 could be somewhere in between these two values, which are represented by the lines.
Some higher metallicity would improve the fit of the red giant branch (RGB) but then the rest of the slope would not be fitted by this model.

\section{Variable stars}
\label{var}

The results of our search for short-period variable stars are given in the~Tab.~\ref{tab:vars}. To find variability we used the Optimal Image Subtraction method, see Alard and
Lupton (1998) and Alard (2000), implemented by Wozniak (2000) and Wojtek Pych\footnote{\texttt{http://users.camk.edu.pl/pych/DIAPL/}}.

We analyzed results of the image subtraction for data from only the two best nights. We were able to detect a few dozen variability events but with such a small amount of data we could not verify most of them. Thus, we report here only the most probable variables, see Tab.~\ref{tab:vars}.

\begin{table}[h]\footnotesize
\begin{center}
\caption{Coordinates and light curve parameters for variable stars of NGC~2155. Periods are given in days. $\rm{A_I}$ is the amplitude of the light curve in the \textit{I}-band.}
\begin{tabular}{ c c c c c c c c}
\hline
Name 		& $\alpha$ & $\delta$ & $\langle$V$\rangle$ & $\langle$V-I$\rangle$	 & $\rm{A_I}$	& Period	& Type \\
\hline
NGC2155-V1  & 5.975308 & -65.467840 & 	20.27			& 0.24					&	0.37	 & 0.05977 & SX Phe \\
NGC2155-V2  & 5.975305 & -65.479066 &  	19.39			& 0.21					&	0.51	 & 0.08949 & SX Phe \\
NGC2155-V3  & 5.976201 & -65.474401 &  	19.41			& 0.46 					&	0.40	 & 1.36083 & Ecl. binary \\
NGC2155-V4  & 5.976681 & -65.463650 &  	20.61			& 1.09 					&	0.26	 & 4.83398 & LP \\
NGC2155-V5  & 5.978492 & -65.471589 & 	18.02			& 1.21 					&	0.06	 & 0.71841 & Red giant \\
NGC2155-V6  & 5.973289 & -65.470010 &  	17.69			& 1.26 					&	0.03	 & 0.34361 & Red giant \\
NGC2155-V7  & 5.976252 & -65.476893 &  	16.98			& 1.39 					&	0.04	 & 0.37629 & Red giant \\
\end{tabular}
\label{tab:vars}
\end{center}
\end{table}

\begin{figure}[h]
\begin{center}
\includegraphics[width=0.8\textwidth, angle=270]{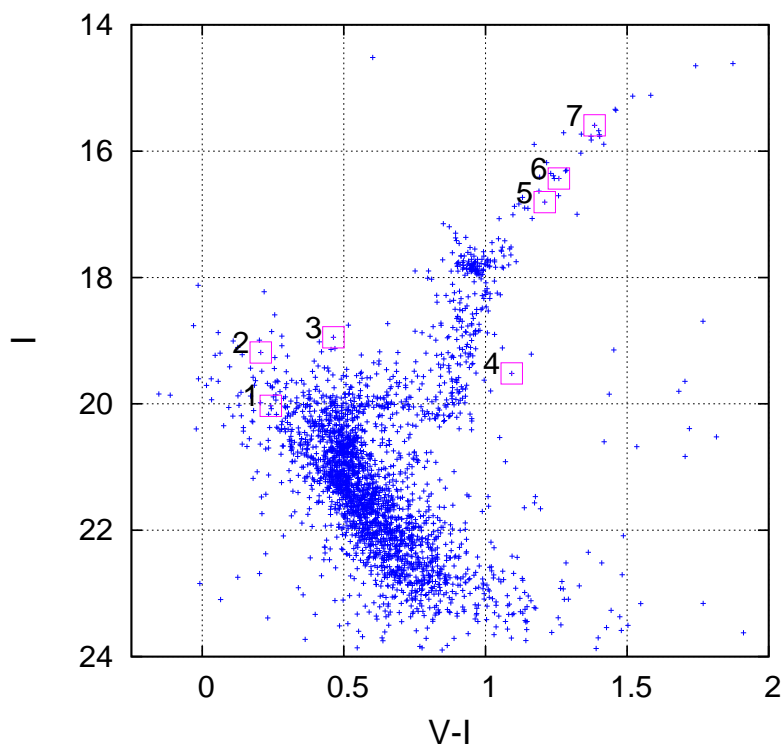}
\caption{The color-magnitude diagram of NGC2155 with marked positions of our variables. Numbers correspond to the last letter of a variable's name given in the~Tab.\ref{tab:vars}}
\label{fig:cmd_var}
\end{center}
\end{figure}

\begin{figure}[htb]
\fbox{\includegraphics[angle=270,scale=0.24]{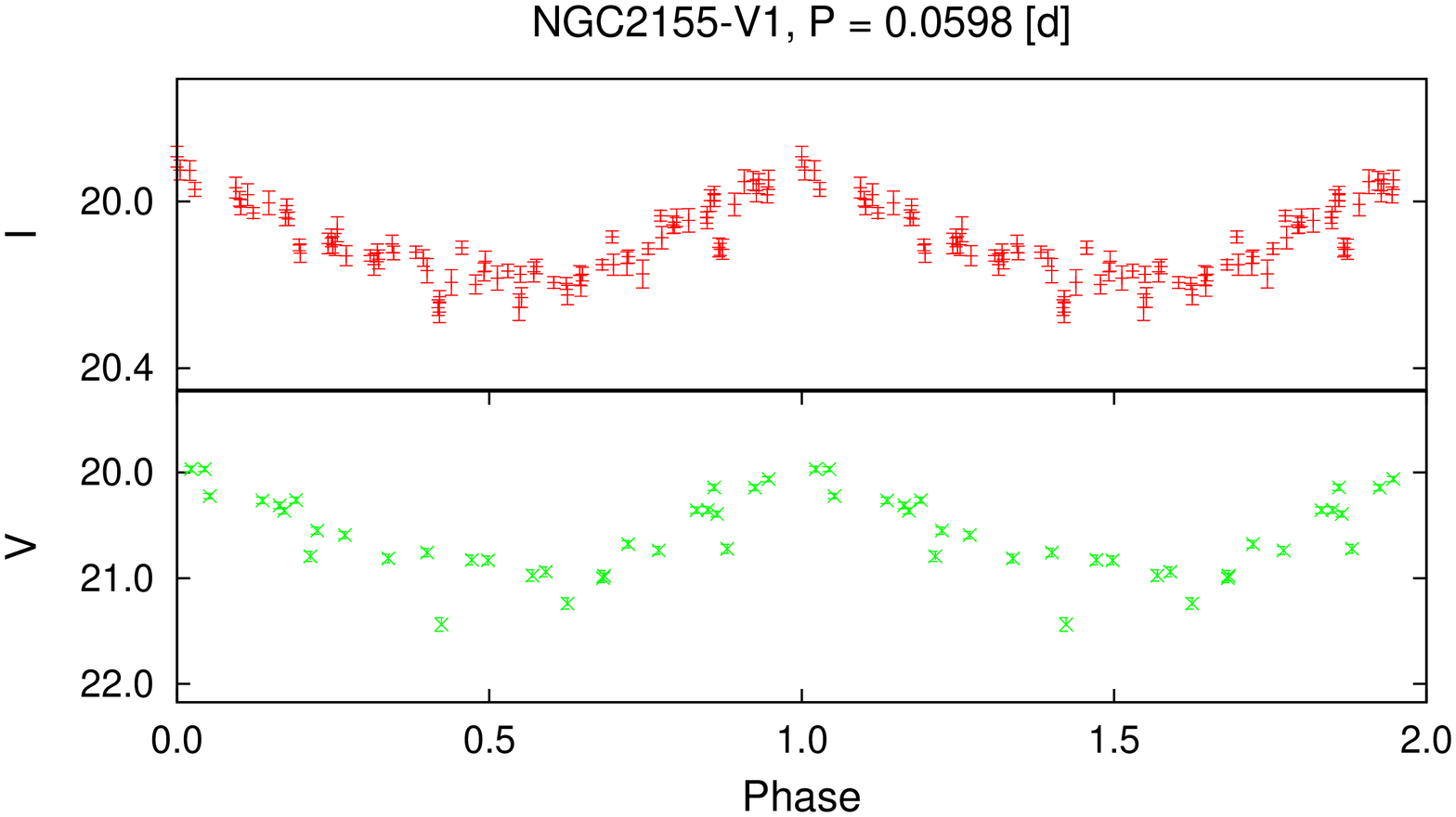}}
\fbox{\includegraphics[angle=270,scale=0.24]{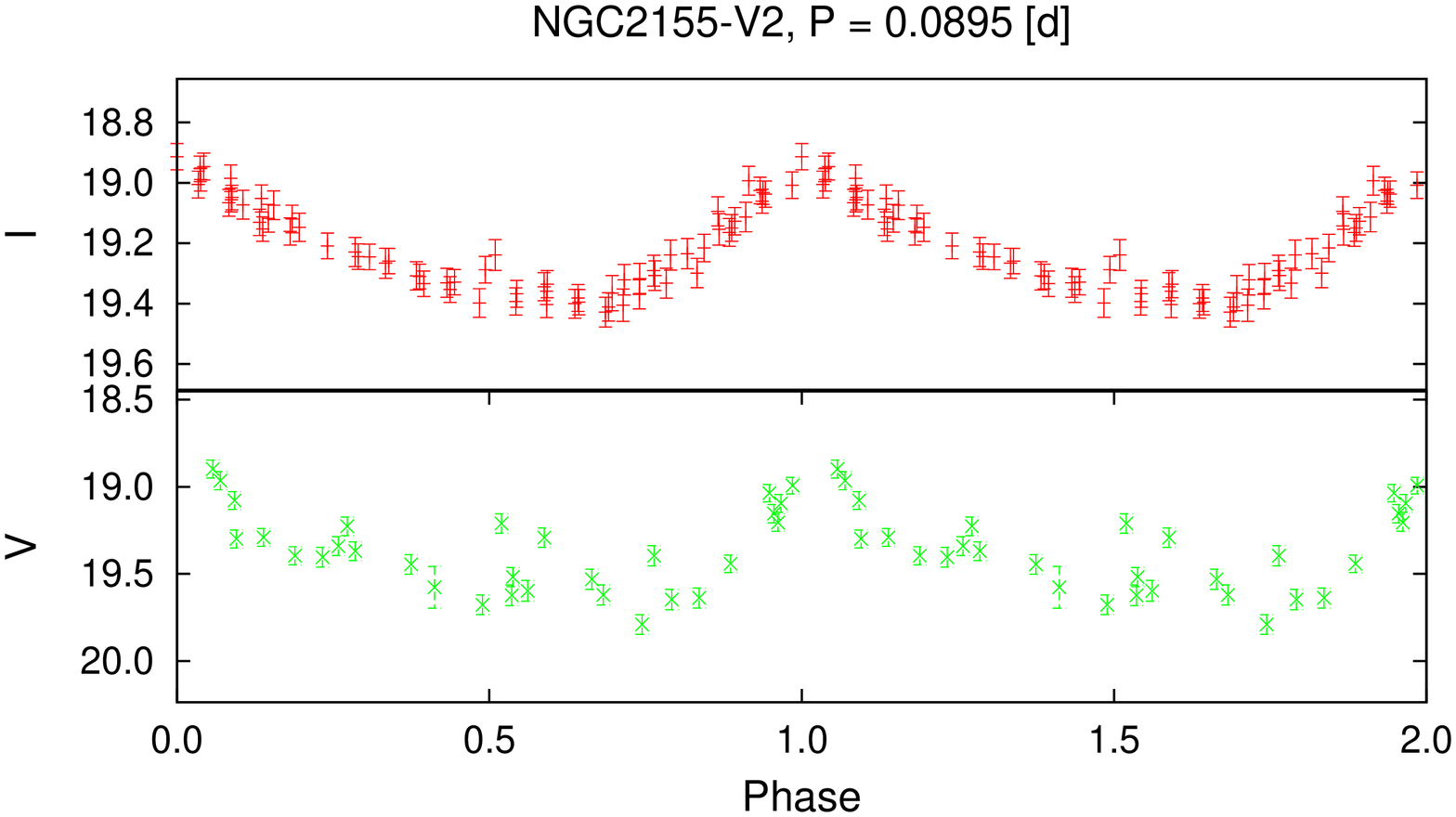}}
\fbox{\includegraphics[angle=270,scale=0.24]{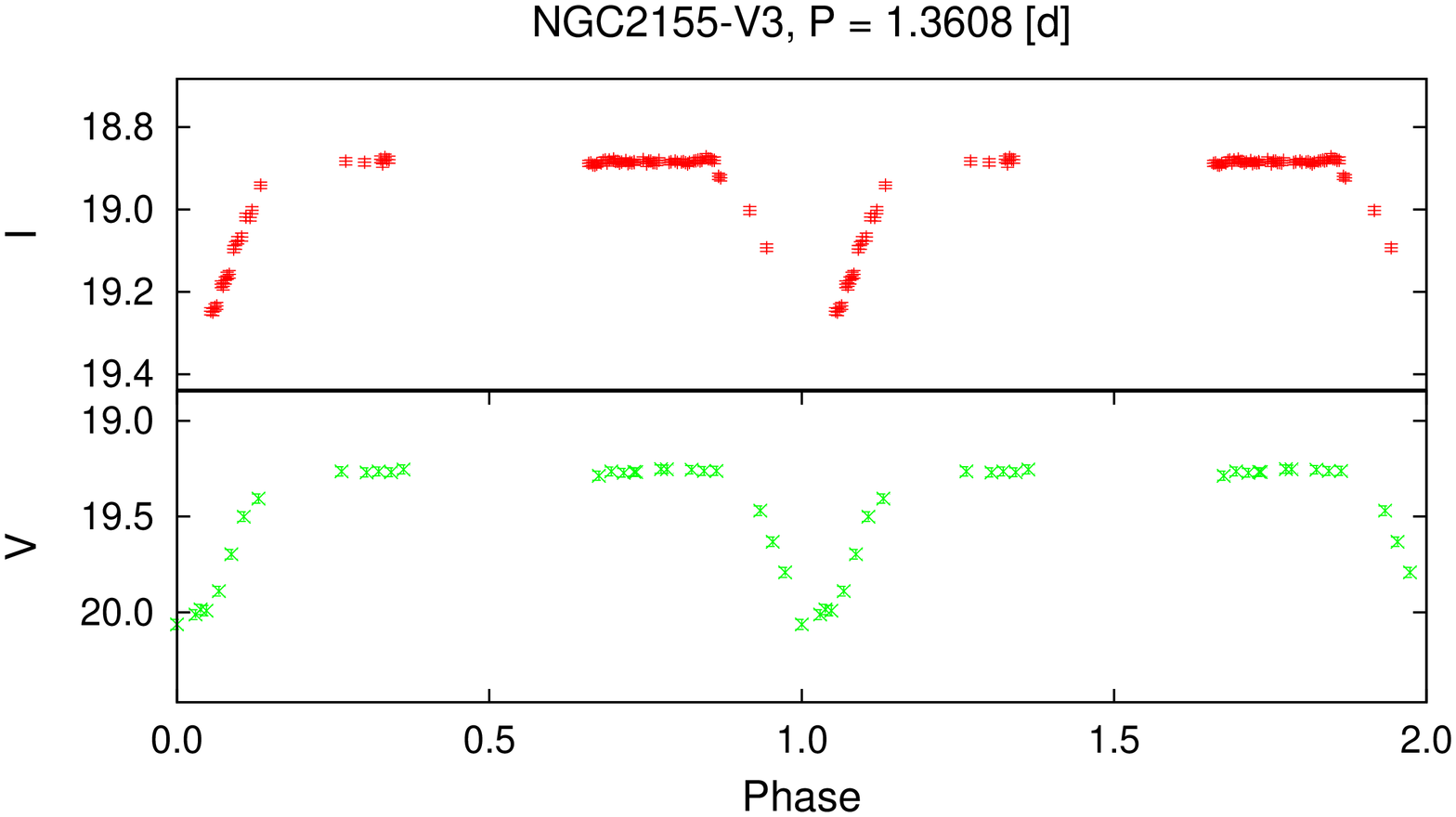}}
\fbox{\includegraphics[angle=270,scale=0.24]{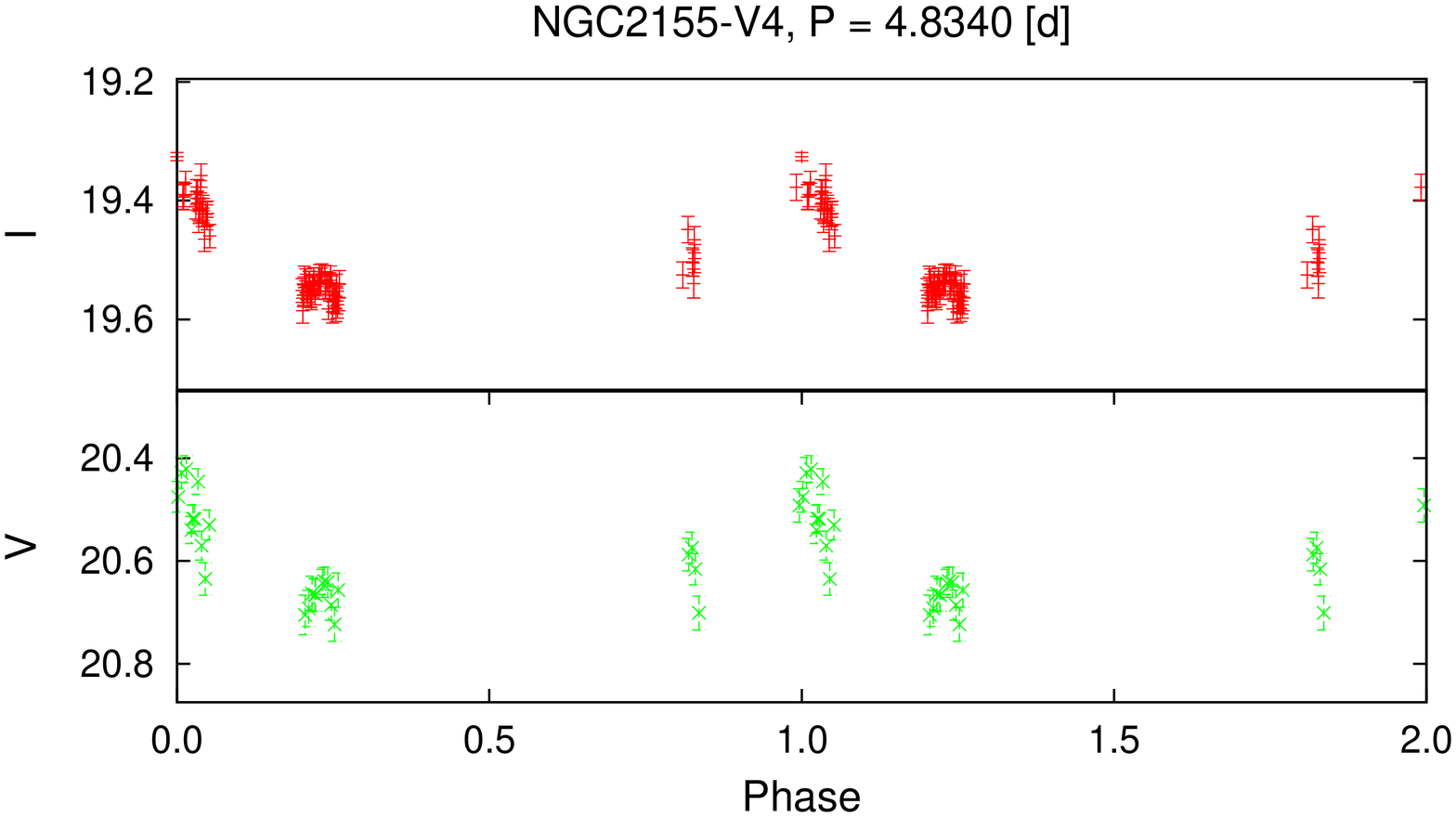}}
\fbox{\includegraphics[angle=270,scale=0.24]{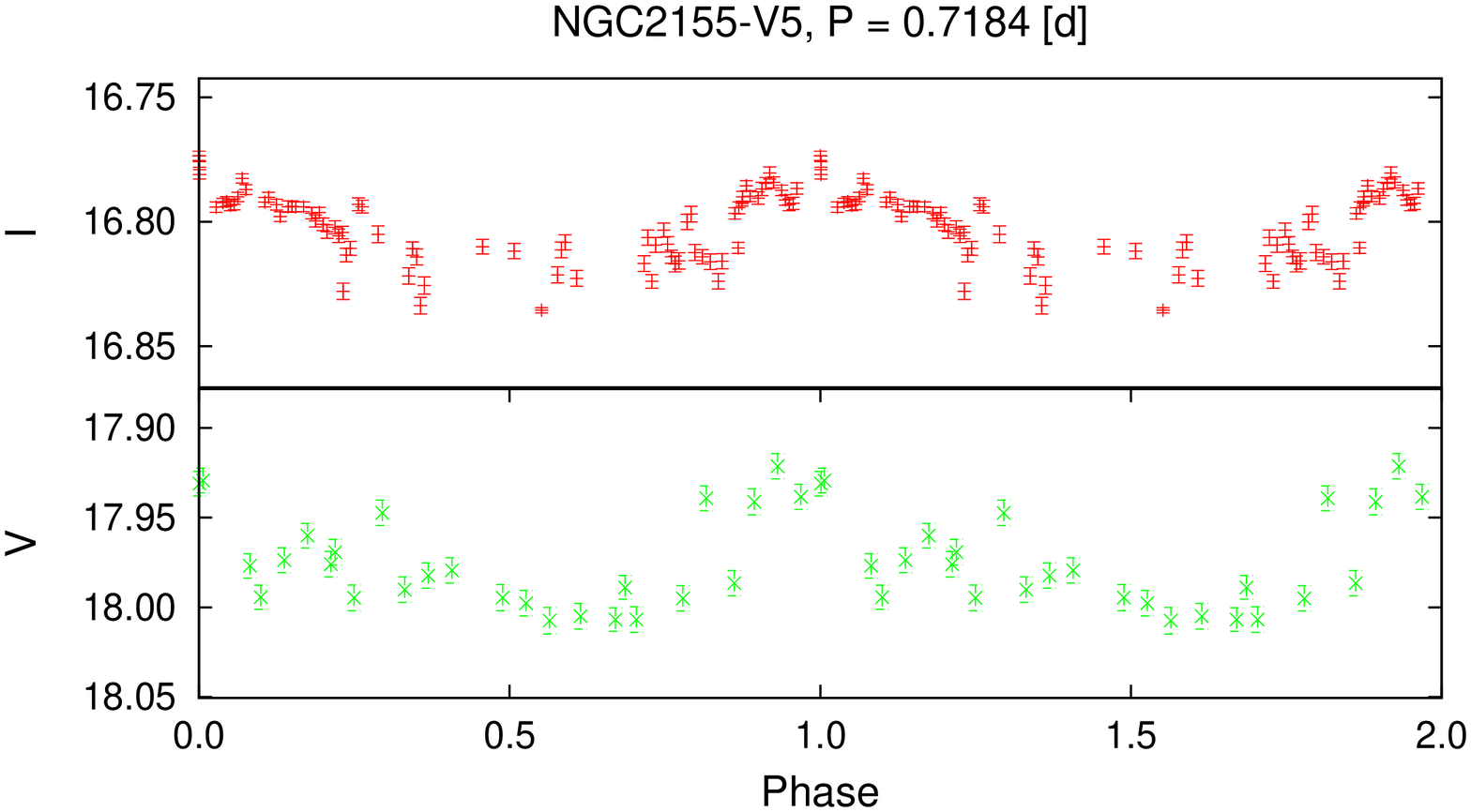}}
\fbox{\includegraphics[angle=270,scale=0.24]{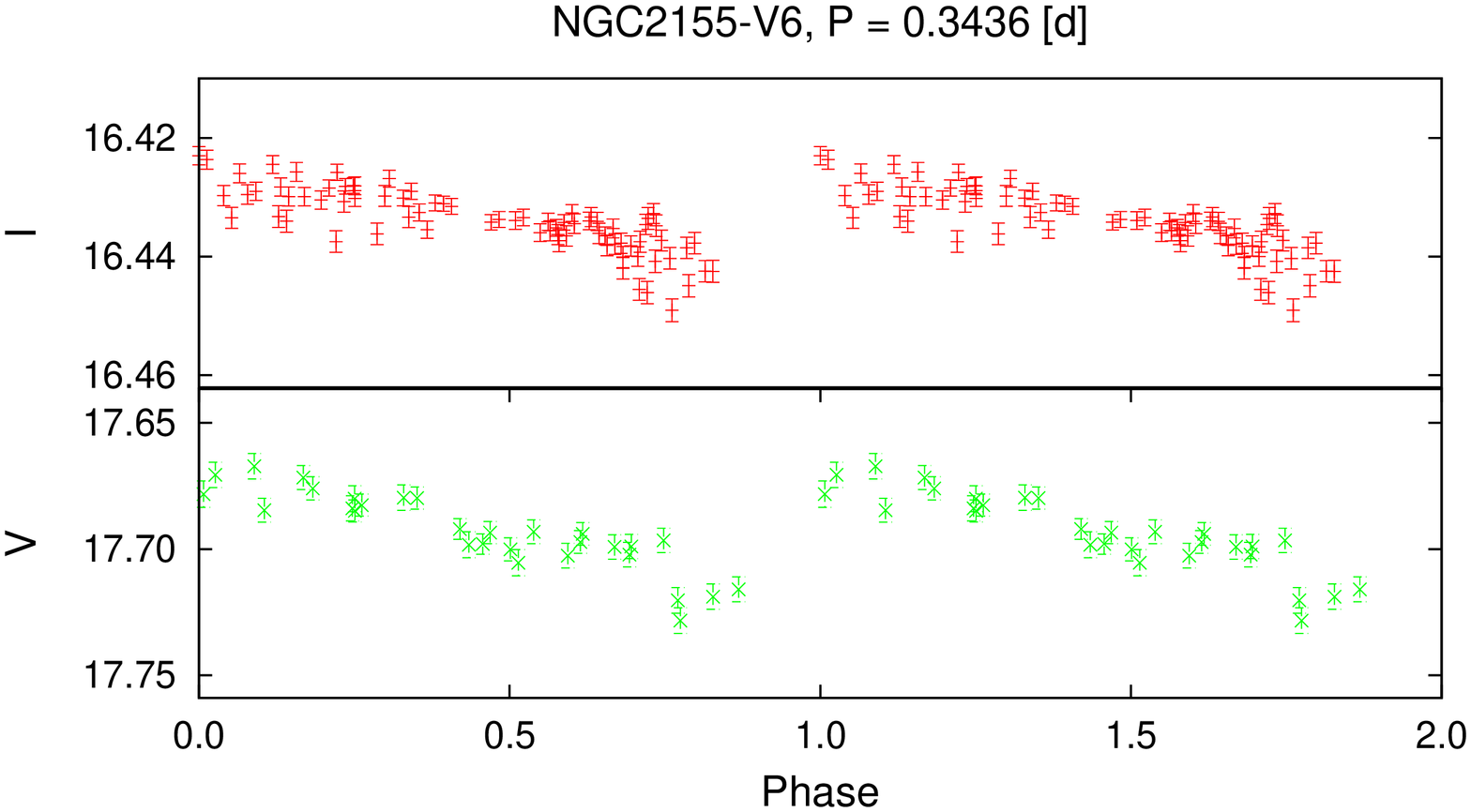}}
\fbox{\includegraphics[angle=270,scale=0.24]{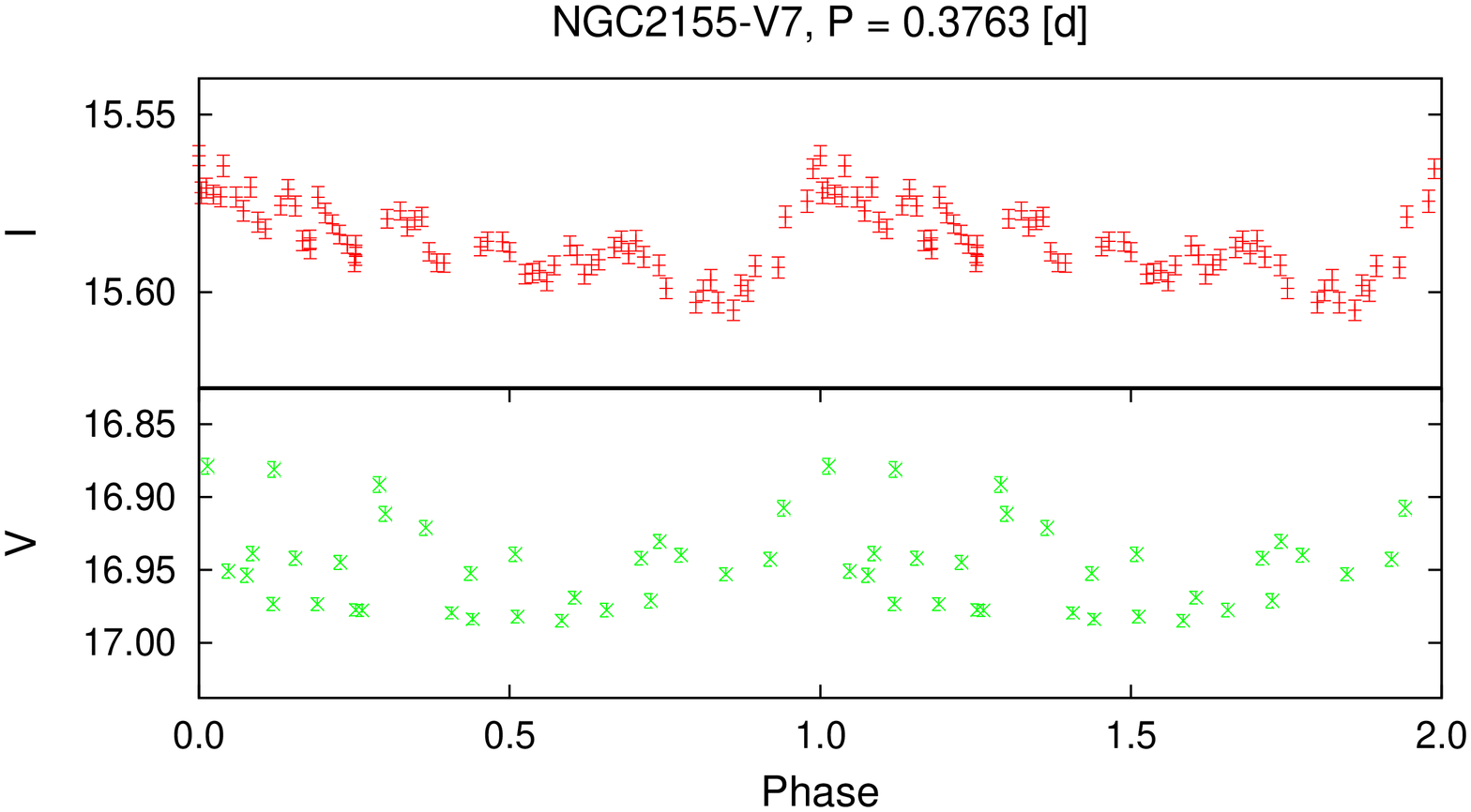}}
\caption{Phased light curves for our newly discovered NGC~2155 variables.}
\label{fig:lc}
\end{figure}

We discovered seven new periodic variables. We were not aware of any previously known variable in NGC~2155. The periods were obtained with the method of Schwarzenberg-Czerny (1996).
The CMD for NGC~2155 with positions of variables is shown in the~Fig.~\ref{fig:cmd_var} and the phased light curves of these stars are given in the~Fig.~\ref{fig:lc}.

The first two variables, \textbf{NGC2155-V1} and \textbf{NGC2155-V2}, can be securely classified as SX~Phe stars. The short periods of 0.059771 and 0.089495~days, 
the trends of their light curves and positions on the CMD in the blue stragglers region are a clear argument for the SX~Phe assigment.
We claim that these stars are pulsating in the fundamental mode, because of their asymmetric light curves and very large amplitudes. The 
 argument that SX~Phe stars with large amplitudes are pulsating in the fundamental mode, and those with $\triangle~V~\leqslant~0.2$~mag in the first overtone was presented for example by Rodr\`iguez and L\'opez-Gonz\'alez (2000) and Santolamazza~et~al. (2001).
Since these stars are more interesting than the remaining variables, we dedicate a separate Section~\ref{sxphe} to them.

The third variable, \textbf{NGC2155-V3}, is a beautiful example of a detached eclipsing binary star. The period given here (1.360828~days) is plausible  but not a certain one
because we did not detect the second eclipse with our short time interval data. This star is 
an attractive target for future observations because such a detached binary can be used to determine the distance to the cluster and to the LMC.

\textbf{NGC2155-V4} is a long period (LP) subgiant star. The light curve here reveals only its
variability; again, the period of 4.833980~days is uncertain. 
The position of NGC2155-V4 on the CMD confirms that this is a LP-type star.

The three remaining variable stars, \textbf{NGC2155-V5}, \textbf{NGC2155-V6} and \textbf{NGC2155-V7}, are classified as red giants, based on their location on the CMD. Their variability is most likely connected with pulsations or/and chromospheric activity.

\section{SX Phe variables}
\label{sxphe}

\subsection{Distance moduli from period-luminosity relations}

SX~Phe variables are pulsating stars of Population~II, very similar to 
$\delta$~Scuti stars which are objects of Population~I. 
Taking into account relatively large distance and crowding in the observed
field we can detect only 
one or two radial modes excited at short periods of $\sim$hours.
In globular clusters SX~Phe stars are
most often found in the blue stragglers stars (BSS) region of the CMD. These
objects are very interesting because their evolution cannot be explained
with the standard theory for a single star. BSS were the subject of many
studies and debates and nowadays it is believed that they are products
of mergers of MS stars or binaries in dense star clusters. Their
existence has an important influence on the evolution of the whole
system, as has been shown for instance by Lombardi and Rasio (2002).

In this context the globular cluster NGC~2155 is interesting for at least
two reasons. First, it consists of a significant population of BSS as can be
seen in our Fig.~6. Second, there are two SX~Phe variables (the detected here
NGC2155-V1 and NGC2155-V2) located in the blue straggler region of the
cluster. If these two variables are real members of the cluster, their
pulsational properties might give strong constrains on the physical 
parameters of the cluster itself. One should remember that, compared
to the Galactic globular clusters, the NGC~2155 is relatively young.
Thus its turn-off point and blue stragglers are much bluer than in
clusters in our Galaxy. A quick comparison with
globular cluster of similar metallicity, 47~Tuc, which age is estimated to around 11.25~Gyr (Thompson~et~al. 2010), 
shows that the turn-off point of NGC~2155 is about 0.3~mag bluer
(see $V/V-I$ CMD of 47~Tuc in Rosenberg~et~al. 2000). 
This might suggest that SX~Phe stars in NGC~2155 are located outside of
the classical pulsation instability strip and their basic properties
(chemical composition?) are different than in Galactic SX~Phe stars.

For our SX~Phe new detections we can derive the distance.
Such stars pulsating in the fundamental mode can be used as "standard
candles" (see references in the Tab.~4). 
We know the periods of these objects; from absolute magnitude - period 
relations we can compute the absolute magnitudes of our SX~Phe stars (see Tab.~4), and from them, the distance modulus.
However, it is challenging to compute it for a longer period SX~Phe, because most of the $M_V - P$ relations are derived from stars with short periods. 
What is more, $M_V - P$ relations depend strongly on the selection criteria 
of the stars which were used to derive these relations. 
It is sometimes challenging to identify the mode of SX~Phe star securely, this leads to a wrong classification of SX~Phe stars, which causes high uncertainties in these relations.

Due to the fact that NGC2155-V1 and NGC2155-V2 have slightly asymmetric
light curves and relatively large peak-to-peak amplitudes of variability, 
we believe that both stars pulsate in the fundamental mode. 
Thus, we can use the known period-luminosity relations to determine their 
absolute magnitudes and the distance modulus. 
Yet, following Gilliland~et~al. (1998), we present several period-luminosity 
(P-L) relations for BSS which have been published in recent years and we
apply them to our data (Tab.~4).

The P-L relation has the following form:
\[  M_V = A \cdot log P + B  \] \\ 
where $A$ and $B$ are constants determined from a set of observed
periods ($P$) and absolute magnitudes ($M_V$) for a sample of stars.

\begin{footnotesize}
\ctable[
caption = Parameters of NGC~2155 V1 and V2 based on various period - absolute magnitude calibrations.
]{ccccccc}{
} {	\FL
A		  	& B		 		&  $\rm{M_{V1}}$	&  $\rm{M_{V2}}$		&	$\rm{(m-M_{V1})_V}$ 	&	$\rm{(m-M_{V2})_V}$ 	&	 reference		\ML
-2.56 		&    0.13		&	3.26		&	2.81			& 	16.84			&	16.41			& 	Nemec, Nemec, and Lutz (1994)	\NN	
-3.29 		&    -1.16		&	2.86		&	2.29			& 	17.24			&	16.94			& 	D. H. McNamara (1995)	\NN	
-3.74 		&    -1.91		&	2.66		&	2.01			& 	17.44			&	17.21			& 	Hog and Petersen (1997)	\NN	
-2.90 		&    -1.20		&	2.35		&	1.84			& 	17.75			&	17.38			& 	Fernie (1992)	\NN	
-3.05 		&    -1.32		&	2.41		&	1.88			& 	17.69			&	17.35			& 	Santolamazza et al. (2001) \NN	
-3.65 		&    -1.38		&	3.08		&	2.45			& 	17.02			&	16.78			& 	Poretti et al. (2008)	\NN	
-3.725 		&	 -1.933		&	2.62		&	1.97			& 	17.48			&	17.25			& 	D. McNamara (1997) \NN
-2.88		&	-0.77		&	2.75		&	2.25			&	17.35			&	16.98			&	Pych et al. (2001)
\LL
}
\end{footnotesize}

The distance moduli for both our SX~Phe variables appear to indicate that
they are foreground stars located in the halo of our Galaxy or perhaps in
remnants of tidal streams of LMC, but not in the cluster. 
On the other hand, probability that
two SX~Phe stars are located exactly half-way between us and LMC and are
additionally placed in the BSS region 
of NGC~2155 must be exceedingly
low. This would be an extraordinary coincidence. 
Thus, in our opinion, these stars may belong to the cluster but their
physical properties may be different than in typical SX~Phe stars in
the Galaxy. We will discuss it in details in the subsection~5.2.

In the literature, 
unsuccessful cases of application of P-L relation
have been described before. 
For instance, Gilliland~et~al. (1998)
found that for SX~Phe stars in 47~Tuc 
the observed magnitudes 
are brighter than calculated from
these relations. This discrepancy can be caused by the metallicity of
these clusters. NGC~2155 and 47~Tuc have approximately the same metallicity: 
[Fe/H]~$\approx$~$-0.71$ and [Fe/H]~$\approx$~$-0.83$ (VandenBerg 2000).
Several authors have already
shown that the pulsation properties of SX~Phe stars are affected by
their metal content, for example
D. McNamara (1997), Rodr\`iguez and L\'opez-Gonz\'alez (2000) and Santolamazza~et~al. (2001).
The studies show that the periods of SX~Phe variables are larger for the clusters 
which have a higher metallicity.

The problem of P-L relations application for  NGC2155-V1 and  NGC2155-V2
is evident. There is of course some uncertainty which comes from these
relations, but the discrepancy in our case is extremely high. To check
whether we deal with some peculiar stars we decided to make further and
more detailed study.

\subsection{Observations vs. theory}

All the calibrations up to obtaining absolute magnitudes $M_V$ and
colors were already done and described in the~Sect.~2. They are valid for
stars in the field of NGC~2155. However, if we want to compare the
observations with models we need to estimate effective temperatures ($T_{\rm{eff}}$) and
luminosities ($L$), i.e. bolometric corrections, for our two SX~Phe stars.
This was done based on Kurucz model atmospheres used in Bessell, Castelli, and Plez 1998 work. 
The resulting values are as follows (errors in luminosity come from the error of distance modulus of NGC~2155):

\begin{table}[h]
\begin{center}
\begin{tabular}{ c | c | c }
Object				&	$\rm{log~T_{eff}}$			& $\rm{log~L}$	  \\
\hline
NGC2155-V1			&		$3.9039 \pm 0.0474$		&	$1.242 \pm 0.052$			\\
NGC2155-V2			&		$3.9125 \pm 0.0456$		&	$1.598 \pm 0.052$
\end{tabular}
\end{center}
\end{table}

\begin{figure}[h]
\begin{center}
\includegraphics[width=0.85\textwidth]{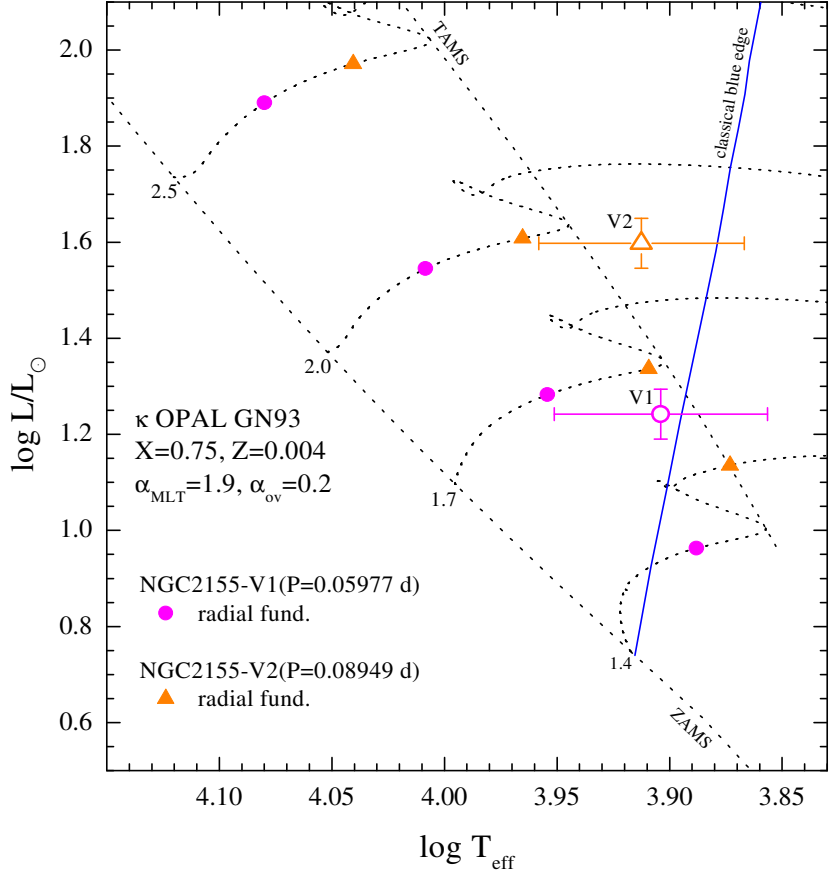} 
\caption{The observed positions of our two SX~Phe stars, NGC2155-V1 and NGC2155-V2, on the HR diagram, plotted with the theoretical main sequence of a few evolutionary tracks for selected masses, which are calculated with the Warsaw-New Jersey code. See details in the text.}
\label{fig:SXPhe}
\end{center}
\end{figure}

Fig.~9 shows theoretical main sequence with a few evolutionary tracks for selected masses,
also with the ZAMS and TAMS (dotted lines). Thick solid line marks the blue edge
of pulsational instability for radial fundamental mode.
Along the selected evolutionary tracks models with fixed periods of radial fundamental mode are marked:
0.05977~days (filled circles) and 0.08949~days (filled triangles) which correspond to observed periods
of variables NGC2155-V1 and NGC2155-V2, respectively. The observed positions of both variables are shown with their
error crosses.

Evolutionary calculations were performed with the Warsaw-New Jersey code using the OPAL opacities
(Iglesias and Rogers 1996) and assuming solar proportions in abundances of chemical elements heavier
than helium according to Grevesse and Noels (1993). The computations were performed for
the hydrogen mass fraction $X~=~0.75$ and heavy element mass fraction (metallicity) $Z~=~0.004$.
Such a choice of chemical composition was caused by the requirement to achieve agreement with
the best fit isochrone.
In stellar envelope, the standard mixing-length theory of convection with the mixing-length
parameter $\alpha (MLT) = 1.9 $(solar-like) was used. For overshooting a two-parameter description
was used (Dziembowski and Pamyatnykh 2008) with the choice of parameters which corresponds
approximately to usual one-parameter description with $ \alpha (ov) = 0.2$.

Linear nonadiabatic analysis of the radial oscillations was performed using a code developed by W.~A.~Dziembowski (for general description see Dziembowski 1977).

It can be seen very clearly that the observed periods do not fit the theoretical values
for standard evolutionary models. Moreover, the stars at positions of observed variables in the HR~diagram
must not be pulsating - they are located left to the blue edge of instability region.

To solve both these problems (fitting the periods as those of the radial fundamental mode,
and ensuring the pulsational instability of this mode), the theoretical models must be
changed both in global stellar mass and chemical composition. 
To fit periods, masses of both variables must be significantly higher than those of standard evolutionary models.
To ensure the instability, it is necessary to increase the helium abundance in the models
(pulsations in the classical instability strip are driven mainly in the second helium ionization
zone in the stellar envelope, so the higher helium abundance results in a widening
of the instability region in the HR~diagram).

The preliminary tests show that it is possible to fit periods and achieve instability at the positions
of variables in the HR~diagram if their stellar masses are by factor $1.8~-~2.3$ larger than evolutionary
values (for NGC2155-V1, the required mass is $2.95~{M}_\odot$, 
$1.8$ times larger of the evolutionary value $1.6~{M}_\odot$; 
for NGC2155-V2, the required mass is $4.1~{M}_\odot$, 
$2.1~-~2.3$ times larger of the evolutionary value $1.8~-~1.9~{M}_\odot$).
To ensure instability for radial fundamental mode, the helium mass fraction, $Y$, must be of about $0.40~-~0.45$
instead of the standard value of about $0.25~(Y=1-X-Z)$. Potentially, such helium-rich stars can be
produced during evolution of close binary systems as result of very effective mass transfer or merger of
the components (we note the paper of Fedorova, Tutukov, and Yungelson (2004), 
especially Figs.~3 and 4 there, which may argue in favor of such a hypothesis). 
We plan to study this hypothesis in more detail.

\section{Summary}
\label{sum}

Using the 6.5m Magellan-1 telescope we obtained  
a Color-Magnitude Diagram for NGC~2155 which
is the deepest one in $I$ and $V$ bands ever published for this cluster. 
Our analysis showed that the age of NGC~2155 is 2.25~Gyr which is even one of the lowest values found for this cluster. Therefore we confirm the location of this cluster outside the age gap of LMC star clusters. NGC~2155 seems to be among the very first clusters formed after the gap, so its age is critical to determine the lower limit of the age gap. 
We found the metallicity of NGC~2155 is [Fe/H]~=~$-0.71$, which is in agreement with previous studies for this cluster.

An interesting results of our study is the discovery of seven variable stars. Among them, three deserve a special attention. 
First, the detached binary star (NGC2155-V3) which should be considered as a target for future studies, because this object could allow us to compute the accurate distance to the NGC~2155. 
Next, two SX~Phe stars (NGC2155-V1 and NGC2155-V2) pulsating in the fundamental mode, with very peculiar behaviour, namely long periods (0.05977 and 0.08949~d, respectively) which seem to be in disagreement with known P-L relations for BSS. 
They cannot be explained by standard evolutionary models and need a further investigation.

\Acknow{
The observing time at the Las Campanas Observatory which was
used for this investigation resulted from an agreement between 
the Carnegie Institution for Science and University of Toronto
on the use of the Magellan Telescopes.

We would like to thank Felipe Sanchez and Hector Nunez: the night assistants at the Baade telescope during the observations. 

This work was supported by the MNiSW grant no.~N~N203~301~335 to AO. 
AAP and TZ acknowledge financial support from MNiSW grant no.~N~N203~379~636. Research of SMR is supported by a grant
from the Natural Sciences and Engineering Council of Canada.
}

\end{document}